\documentclass[amsmath, amsfonts, superscriptaddress, showpacs, twocolumn, prl]{revtex4}
\usepackage{graphicx}
\usepackage{epsfig}
\usepackage{bm}
\usepackage{dcolumn}
\usepackage{amsmath}
\usepackage{amssymb}
\usepackage[varg]{txfonts}

\newcommand{\Det}{\mathop{\rm det}\nolimits}

\begin{document}

\title{Nonequilibrium noise and current fluctuations at the superconducting phase transition}

\author{Dmitry Bagrets}
\affiliation{Institut f\"{u}r Theoretische Physik, Universit\"{a}t
zu K\"{o}ln, Z\"{u}lpicher Str. 77, 50937 K\"{o}ln, Germany}

\author{Alex Levchenko}
\affiliation{Department of Physics and Astronomy, Michigan State
University, East Lansing, Michigan 48824, USA}

\begin{abstract}
We study non-Gaussian out-of-equilibrium current fluctuations in a mesoscopic NSN circuit at the point of a superconducting phase transition. The setup consists of a voltage-biased thin film nanobridge superconductor (S) connected to two normal-metal (N) leads by tunnel junctions. We find that above a critical temperature fluctuations of the superconducting order parameter associated with the preformed Cooper pairs mediate inelastic electron scattering that promotes strong current fluctuations. Though the conductance is suppressed due to the depletion of the quasiparticle density of states, higher cumulants of current fluctuations are parametrically enhanced. We identify an experimentally relevant transport regime where excess current noise may reach or even exceed the level of thermal noise.
\end{abstract}

\date{November 10, 2014}


\maketitle

\textit{Introduction}.-- Fluctuations of the order parameter
associated with preformed Cooper pairs strongly influence the transport
properties of superconductors above the critical temperature $T_c$.
Owing to extensive research spanning over several decades we
have learned a lot about the thermodynamic and kinetic properties
in the fluctuation regime~\cite{Book}. In the context of
transport, fluctuation-induced corrections to electric, thermal,
thermoelectric, and thermomagnetic kinetic coefficients have been
rigorously established within the linear response formalism. However,
despite its long history, little is known about
the nonlinear~\cite{Dorsey,LO,AL} or nonequilibrium
domains~\cite{Aronov,Kogan,Nagaev}. In particular, the answer to the
question on how superconducting fluctuations affect the
noise or higher-order correlation functions of various observables
remains open. We address this outstanding problem by studying excess
current noise in a system where a superconductor is tailored to be in
the fluctuation regime above $T_c$ and driven out of equilibrium
by an externally applied voltage. Interestingly, this problem has a very
natural connection to another rich field, namely, the full counting
statistics (FCS) of electron transfer~\cite{FCS} in mesoscopic systems. 
It concentrates on finding a probability distribution function for the number of
electrons transferred through the conductor during a given period of
time. FCS yields all moments of the charge transfer, and in general
it encapsulates complete information about electron transport,
including the effects of correlations, entanglement, and also
information about large rare fluctuations. To access the 
FCS experimentally is a challenging task, however, great progress has been achieved 
during the last decade in the field of quantum noise~\cite{Exp-1,Exp-2,Exp-3,Exp-4,Exp-5,Exp-6,Exp-7,Exp-8,
Exp-9,Exp-10,Exp-11,Exp-12,Exp-13}, where new detection schemes have enabled
the extension of traditional shot noise measurements to higher-order current correlators.

This work serves a dual purpose. First, we elucidate the effect of
superconducting fluctuations on the nonequilibrium transport and
derive a cumulant generating function for FCS of current fluctuations
in a mesoscopic proximity circuit that contains, as its element,
a fluctuating superconductor. We find that, due to a depletion of the
quasiparticle density of states, the conductance of the device under
consideration is suppressed, however, noise and higher moments of the
current fluctuations are enhanced due to inelastic electron scattering
in a Cooper channel. It should be stressed that finding the FCS 
for interacting electrons is a very challenging task, with only a
few analytical
results known to date~\cite{Andreev-PRL01,Kindermann-PRL03,Dima-PRL04,Dima-PRL05,
Kindermann-PRL05,Gogolin-PRL05,Gutman-PRL10} (see also the review
articles~\cite{Dima-Review-1,Dima-Review-2}).

The second important aspect of this paper is a derivation of the
nonequilibrium variant of the time-dependent Ginzburg-Landau action
(TDGL). The conventional paradigm behind TDGL
phenomenology~\cite{TDGL} and its subsequent
generalizations~\cite{TDGL-1,TDGL-2,TDGL-3,TDGL-4,TDGL-5,TDGL-6} is to assume
that electronic (quasiparticle) degrees of freedom are at
equilibrium and concentrate on the dynamics of the order parameter
field. While leading to correct static averages,
fluctuation-dissipation relations, and gauge invariance, this way of
handling the problem fails to provide any prescription for
calculating the higher moments of observables, even at equilibrium.
Furthermore, existing theories exclude the stochastic nature
of electron scattering on the order parameter fluctuations.
Technically, the inclusion of such effects should result in stochastic
noise terms (Langevin forces) which have a feedback on
superconducting fluctuations. Below we elaborate on the methodology that
includes all these effects.

\begin{figure}[t]
  \includegraphics[width=7cm]{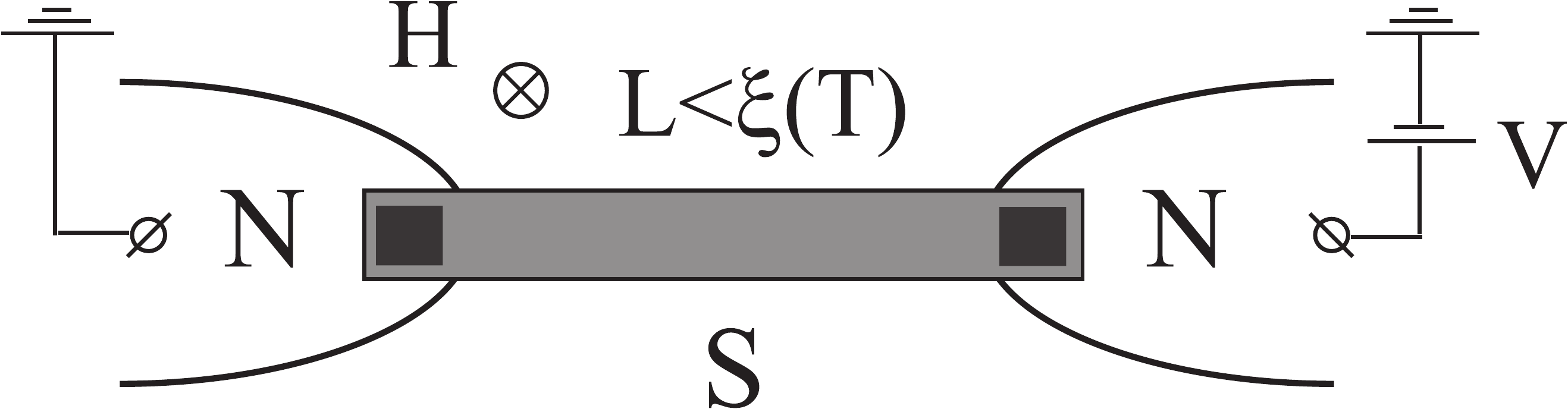}\\
 \caption{The layout of a mesoscopic NISIN proximity circuit
 under voltage bias, magnetic field, and at a temperature 
 above $T_c$ of a superconductor.}\label{Fig-NSN}
\end{figure}

\textit{Model and results}.-- We consider a superconducting diffusive
wire (nanobridge) of length $L$ connected to two normal reservoirs by
tunnel junctions with dimensionless conductances $g_1$ and $g_2$, thus
forming a NISIN-structure (Fig.~\ref{Fig-NSN}). For the conductance of the wire we assume 
$g_W>g_{1,2}$ and, moreover, $g_{1,2}\gg 1$, so that
charging effects can be neglected. The system is driven out of equilibrium by
the finite bias $eV$, and we will
limit ourselves to the regime $T-T_c\lesssim eV\ll T_c$. We also
consider an externally applied magnetic field
$H$, which leads to dephasing of the Cooper pairs due to orbital
effects. We concentrate on the temperature regime in the immediate
vicinity of the critical temperature of a superconductor. In this case, electron transport is dominated by
interaction effects in the Cooper channel, which are singular in
$\Delta T=T-T_c$. Finally, we assume
$L\lesssim\xi(T)\simeq\sqrt{D/(T-T_c)}$, where $\xi$ is a superconducting coherence length 
and $D$ is a diffusion coefficient in the wire. This  assumption greatly simplifies
the problem by making it effectively zero dimensional when neglecting gradient terms in the 
effective low-energy action. We note that such devices are readily available
in experiments~\cite{Exp-14,Exp-15,Chan-PRL05,Chan-PRL09,Chen-PRL09,Chen-PRB11,Aref-PRB12, Li-PRB11,Kamenev_NPhys14} and find their practical implementation 
as superconducting hot electron bolometers~\cite{Prober, Exp-Grenoble}.

Our goal is to derive the cumulant generation function (CGF) ${\cal F}(\chi)$ for the irreducible moments 
of current fluctuations. It is defined as a logarithm of the nonequilibrium partition function, 
${\cal F}(\chi) = -\ln\mathcal{Z}(\chi)$, where the counting field $\chi$ is the variable conjugated 
to the classical part of the current~$I$. Derivatives of ${\cal F}(\chi)$ give the average value of the current,
shot noise, and higher-order moments $C_n$ of charge transfer during a long observation time $t_0$.
 
\begin{figure}[t]
  \includegraphics[height=3.35cm]{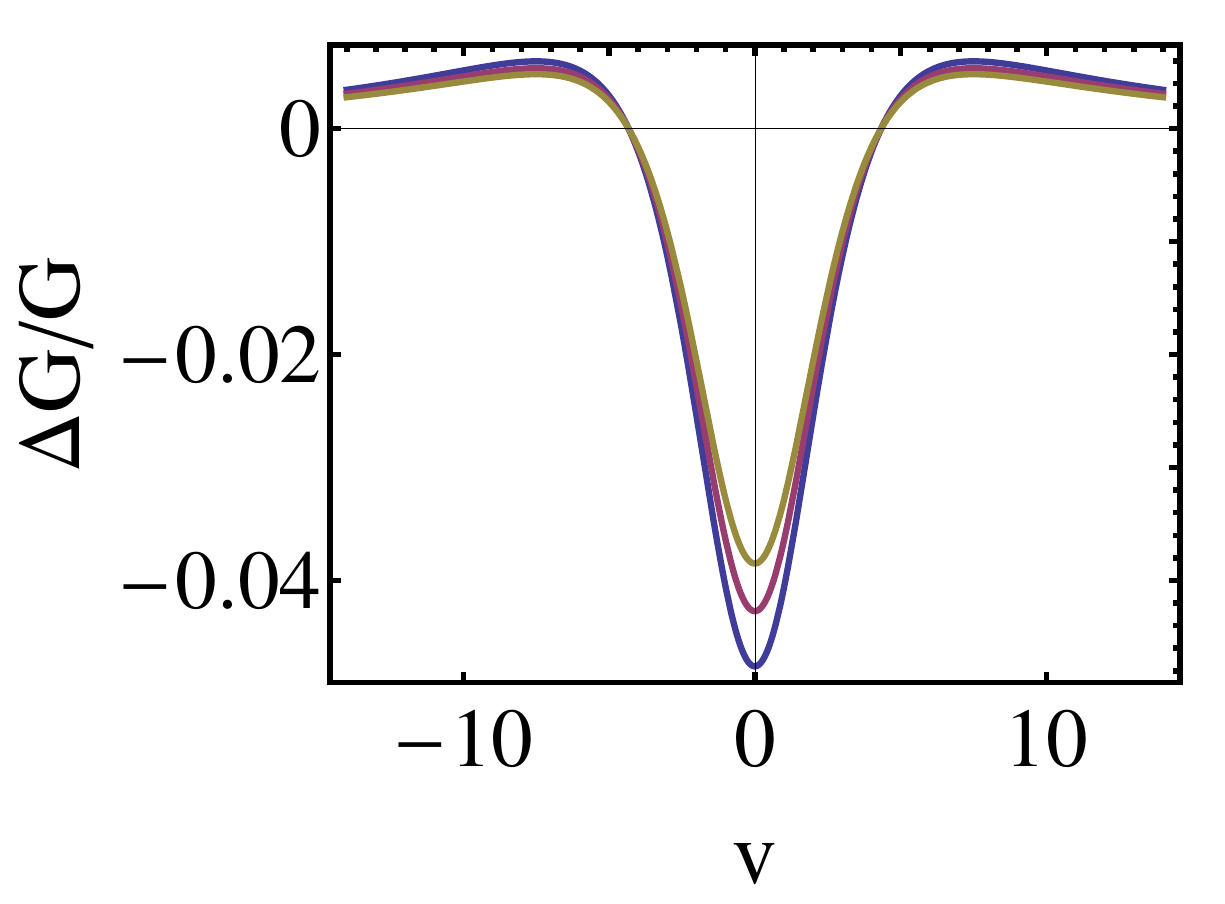}
  \includegraphics[height=3.35cm]{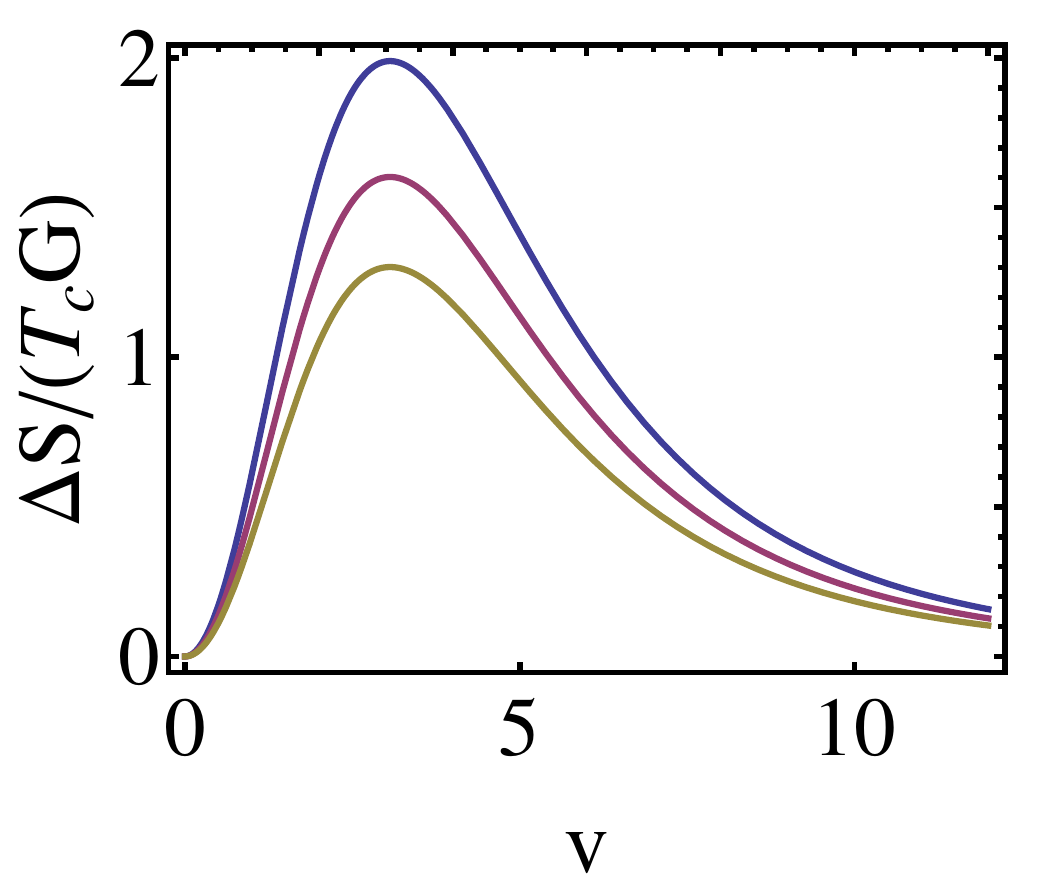}\\
  \caption{Fluctuation-induced correction to conductance $\Delta G$ normalized
  to the normal conductance $G$ (left), and normalized fluctuation-induced nonequilibrium excess current noise in units of thermal noise power $T_cG$ (right),
  plotted vs bias voltage $v=eV/\sqrt{T_c\Delta T}$ for $\Delta T = \delta$,
  $g=2.5\times 10^3$ and $\Gamma/T_c=0.25, 0.5, 0.75$.}\label{Fig2}
\end{figure}

In the normal state away from $T_c$, where superconducting
correlations are negligible, the above device represents a double
tunnel junction. In this case CGF is easy to
compute (see, e.g., Ref.~\cite{Dima-Review-1}). 
The effects of a Coulomb interaction on conductance and
current noise in a similar setup have been previously addressed
on the basis of the quantum kinetic approach
equation~\cite{Ahmadian-PRB05,Catelani-PRB07}.
Superconducting correlations in the vicinity of $T_c$ strongly
affect CGF already at low bias, $eV\ll T_c$. We delegate
a derivation to the end of the paper and first present our main result, 
\begin{equation}\label{lnZ-SC}
{\cal F}(\chi)=-t_0T_c\mathcal{E}
\left[1-\sqrt{1-2\left(\chi^2-\frac{ieV\chi}{T_c}\right)
\frac{\phi+\eta\mathcal{E}}{\mathcal{E}^2}}\right],
\end{equation}
which accounts for inelastic scattering of electrons on
superconducting fluctuations while traversing across the wire. The
proximity to a superconducting transition is controlled by the
function
\begin{equation}
\mathcal{E}(\Delta T,V)=a\frac{\Delta
T}{T_c}+b\frac{(eV)^2}{T^2_c},\quad a=\frac{8}{\pi},\quad
b=\alpha_1\alpha_2\frac{14\zeta(3)}{\pi^3},
\end{equation}
where $\alpha_k=g_k/(g_1+g_2)$ and $\zeta$ is the Riemann zeta
function. At finite magnetic field the critical temperature is
downshifted according to the law
$\ln(T_c/T_{c0})=\psi\big(\frac{1}{2}\big)-\psi\big(\frac{1}{2}+\frac{\Gamma}{4\pi
T_c}\big)$, where $T_{c0}=T_c(H=0)$, $\Gamma=E_{Th}+\tau^{-1}_H$ and
$\psi$ is the digamma function. Thouless energy
$E_{Th}=(g_1+g_2)\delta/4\pi$ is defined through the mean level
spacing in the wire $\delta$, while the dephasing time
$\tau^{-1}_H\simeq (D/L^2)(\Phi/\Phi_0)^2\propto H^2$ is due to
orbital effects of the perpendicular magnetic field, where $\Phi$ is a total magnetic flux
through the wire and $\Phi_0$ is the flux quantum. The two
dimensionless functions in Eq.~\eqref{lnZ-SC} are defined as follows:
\begin{subequations}
\begin{equation}
\phi=-\frac{\alpha_1\alpha_2}{\pi^3}\left(\frac{E_{Th}}{T_c}\right)
\psi''\left(\frac{1}{2}+\frac{\Gamma}{4\pi
T_c}\right),
\end{equation}
\begin{equation}
\eta=\frac{2\alpha_2\alpha_2}{\pi^3}\left[\frac{E_{Th}}{\pi\Gamma}
\psi'''\left(\frac{1}{2}+\frac{\Gamma}{4\pi
T_c}\right)-\psi''\left(\frac{1}{2}+\frac{\Gamma}{4\pi
T_c}\right)\right].
\end{equation}
\end{subequations}
The effect of fluctuations is the most singular provided that
$E_{Th}\gg\Delta T$ where $\phi\gg\eta\mathcal{E}$. In this case
Eq.~\eqref{lnZ-SC} yields a conductance correction
\begin{equation}\label{Conductance}
\frac{\Delta
G}{G_Q}=\frac{2\alpha_1\alpha_2}{\pi^2}\left(\frac{E_{Th}}{\Delta
T}\right)\psi''\left(\frac{1}{2}+\frac{\Gamma}{4\pi
T_c}\right)\frac{a-bv^2}{(a+bv^2)^2},
\end{equation}
where we introduced a notation $v=eV/\sqrt{T_c\Delta T}$. This result
is plotted in Fig.~\ref{Fig2}~(left) for a certain choice of parameters
versus bias voltage and has a BCS-like density of states profile (note
that $\Delta G$ is actually negative since $\psi''<0$). The latter
should not be surprising since superconducting fluctuations deplete
energy states near the Fermi level, which leads to a zero-bias
anomaly. In the same limit we find an excess current noise power,
\begin{equation}\label{Noise}
\frac{\Delta
S_I}{G_QT_c}=\frac{4\alpha^2_1\alpha^2_2}{\pi^5}\left[\psi''\left(\frac{1}{2}+\frac{\Gamma}{4\pi
T_c}\right)\right]^2\!\!\left(\frac{E_{Th}}{\Delta T}\right)^2\!\!
\frac{v^2}{(a+bv^2)^3},
\end{equation}
which is plotted in Fig.~\ref{Fig2}~(right). The low frequency dispersion of
the noise is set by $\omega=\mathrm{max}\{\Delta T,(eV)^2/T_c\}$.
From Eq.~\eqref{lnZ-SC} one can extract the $n^{th}$-moment of the
current fluctuations which progressively display more singular
behavior,
\begin{equation}
\label{eq:In}
C_n=\langle I(\omega_1)\ldots I(\omega_n)\rangle_{\omega_k\to0}\!\simeq
e^{n-2}G_QT_c\!\left(\frac{E_{Th}}{\Delta
T}\right)^n\!\!\left(\frac{T_c}{\Delta T}\right)^{n/2-1}.
\end{equation}
We interpret this result as bunching of electrons due to
slow time-dependent fluctuations of the order parameter, which result in
a long avalanches of charges and thus parametrically enhanced
current fluctuations.

\begin{figure}[t]
  \includegraphics[width=8cm]{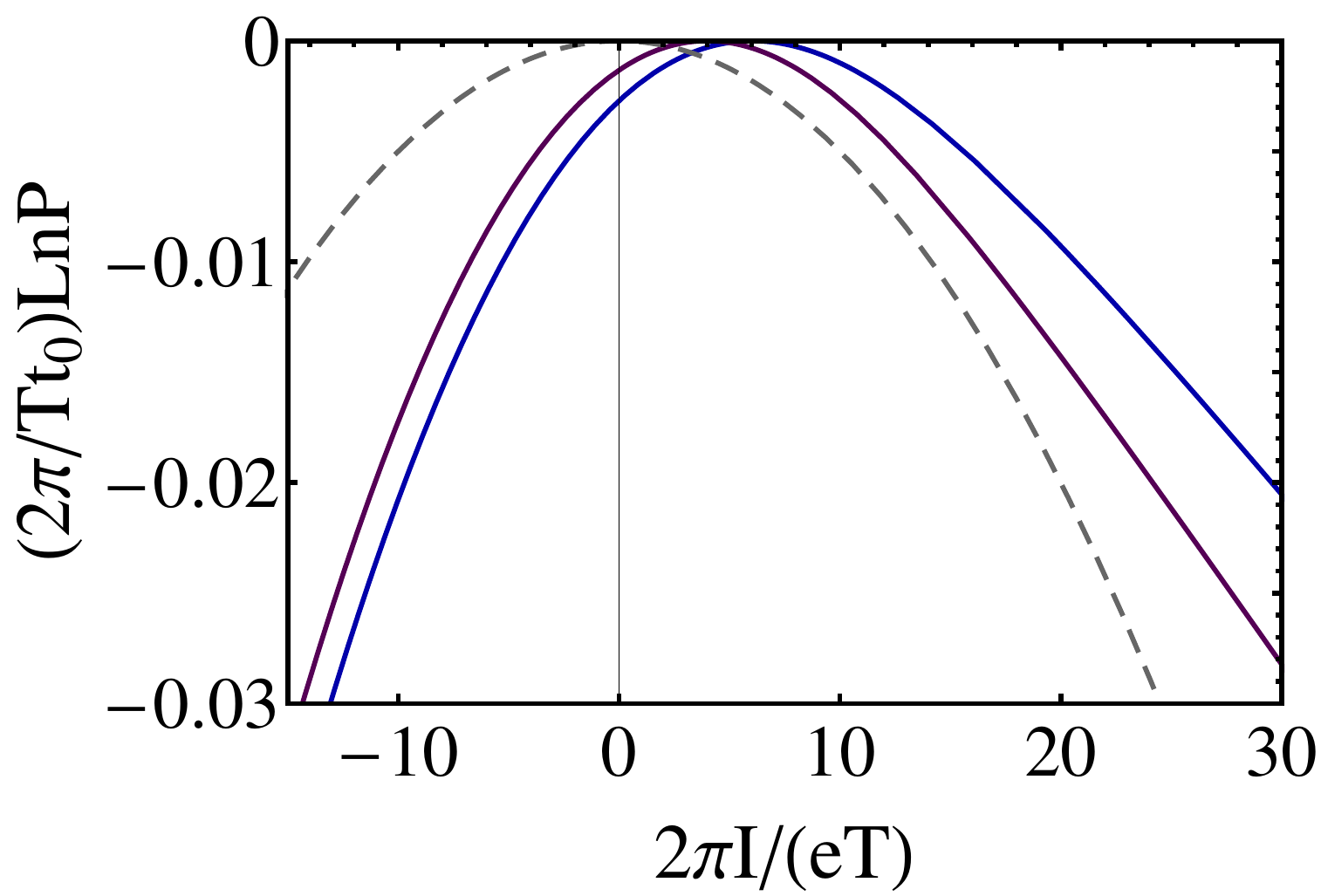}\\
  \caption{The logarithm of probability $P(I)$ to measure current fluctuations plotted vs the current  $I$. The parameters are the same as in Fig.~\ref{Fig2}, with $\Gamma/T_c=0.25$. Top curve: $v=3.0$; middle curve:$v=1.5$; dotted curve: $v=0$ (Gaussian thermal fluctuation).}\label{Fig3}
\end{figure}

This conclusion is substantiated by the direct analysis of the current probability distribution
defined by $P(I)=(2\pi)^{-1}\int_{-\pi}^{\pi} \exp\{-{\cal F}(\chi) + i \chi (It_0/e)\} d\chi$. We estimate 
this integral using the saddle point method by rotating the integration contour to complex $\chi$.
The typical result is plotted in Fig.~\ref{Fig3}. The resulting distribution has a long exponential
tail $P(I) \propto \exp\{ - \lambda (It_0/e) \}$ for positive currents $I$ originating from the branch point 
$\chi = i \lambda$ of the CGF and describing avalanches of transferred charges. In the limit $E_{\rm Th} \gg \Delta T$ one finds $\lambda = (T_c/eV)
(\mathcal{E}^2/2\phi)$, which gives an estimate $\lambda \sim (1/g)(\delta/T_c)^{1/2} \ll 1$ at the direct vicinity
of the phase transition when $\Delta T \sim \delta$ and $v \sim 1$. The latter result is in agreement with
Eq.~(\ref{eq:In}). We note that parametric enhancement of current fluctuations is a universal phenomenon
whenever soft modes are present in the system, and is known to occur, e.g., in interacting diffusive 
mesoscopic wires~\cite{Dima-PRL04} or in molecular junctions~\cite{Utsumi:2013}.

\textit{Estimates}.-- Let us now discuss the experimentally relevant parameters to observe the effect and estimate its actual magnitude. The maximal value of the nonequilibrium excess current noise normalized to the thermal noise at $T_{c0}$ that follows from Eq.~\eqref{Noise} is
\begin{equation}
\left(\frac{\Delta S_{I}}{GT_{c0}}\right)_{\mathrm{max}}\simeq\frac{1}{25g}\left(\frac{E_{Th}}{\Delta T}\right)^2.
\end{equation}
When finding this estimate we took symmetric structure $\alpha_1=\alpha_2=1/2$, used $\psi''(1/2)=-14\zeta(3)$ and assumed $\Gamma/4\pi T_c\ll1$. This condition will be justified below. The minimal allowed $\Delta T$ in our theory is limited by the mean level spacing. Indeed, since $E_{Th}/\Delta T=g/2\pi$ at $\Delta T=\delta$, then the fluctuation-induced correction to conductance $\Delta G$ in Eq.~\eqref{Conductance} already reaches its bare value and thus our approach breaks down for the lower $\Delta T$. At that bound the noise remains parametrically enhanced, $(\Delta S_I/GT_{c0})_\mathrm{max}\simeq g/100\pi^2$, since $g\gg1$, however, a large numerical factor in the denominator significantly diminishes the actual magnitude of the effect. Now we look for realistic numbers. For the layout design in Fig.~\ref{Fig-NSN} we assume a wire of length $L\simeq0.5\mu$m and width $w\simeq100$nm be made of a two-dimensional film of thickness $d\simeq10$nm. For aluminum nanowires 
the typical diffusion coefficient is $D\simeq10^2$cm$^2$s$^{-1}$, the Fermi velocity is $v_F=2\cdot10^8$cm/s, and resistivity $\rho\simeq2\mu\Omega$cm. These numbers provide a Thouless energy $E_{Th}=D/L^2\simeq0.3$K, a mean free path $l=3D/v_F\simeq15$nm, a diffusive coherence length at zero temperature $\xi=\sqrt{\xi_0l}\simeq440$nm, where $\xi_0=v_F/T_{c0}\simeq1.3\mu$m for the bulk aluminum $T_{c0}=1.2$K, and a sheet resistance $\rho_\Box=\rho/d\simeq2\Omega$. The latter translates into the normal wire resistance $R_W=\rho_\Box L/w\simeq10\Omega$ and the dimensionless conductance $g=1/G_QR_W\simeq2.5\cdot10^3$ of the nanostructure.  The corresponding mean level spacing is $\delta=2\pi E_{Th}/g\simeq0.75$mK while $\Gamma/ T_{c0}\simeq 0.25$.  Finally, the realistic estimate for maximal nonequilibrium noise above its thermal level is $(\Delta S_I/GT_{c0})_{\mathrm{max}}\simeq2$, as shown in Fig.~\ref{Fig2} (right). Similar estimates can be carried out for zinc and lead nanowires. All these parameters are within the reach of current nanoscale fabrication technology and high precision measurements. 
  
\textit{Formalism}.-- As a technical tool to derive
Eq.~\eqref{lnZ-SC}, we use the Keldysh technique built into the framework
of the nonlinear-sigma-model (NL$\sigma$M)~\cite{NLSM-1,NLSM-2,NLSM-3}.
For the above specified conditions, separation of the length scales $l\ll \xi(0)\ll L\ll \xi(T)$ implies a diffusive limit and the quantum action of the device
under consideration (Fig.~\eqref{Fig-NSN}) is given by the following
expression $\mathcal{S}=\mathcal{S}_Q+\mathcal{S}_\Delta+\mathcal{S}_T+\mathcal{S}_H$, where
\begin{subequations}
\begin{equation}\label{S-Q}
\mathcal{S}_Q=\frac{i\pi}{\delta}\mathrm{Tr}\big(-\hat{\tau}_3\partial_t\hat{Q}+i\hat{\Delta}\hat{Q}\big),\quad
\mathcal{S}_\Delta=-\frac{2}{\lambda\delta}\mathrm{Tr}\big(\vec{\Delta}^\dag\hat{\sigma}_1\vec{\Delta}\big),
\end{equation}
\begin{equation}\label{S-T}
\mathcal{S}_T=\frac{i}{16}\sum_{k=1,2}g_k\mathrm{Tr}\big\{\hat{Q}^{[\chi]}_k,\hat{Q}\big\},\quad
\mathcal{S}_H=\frac{i\pi}{8\delta\tau_H}\mathrm{Tr}\big(\hat{\tau}_3\hat{Q}\big)^2.
\end{equation}
\end{subequations}
Here $\delta$ is the mean level spacing in the island, and $\lambda$ is
the coupling constant in the Cooper channel. The two sets of Pauli
matrices $\hat{\tau}_i$ and $\hat{\sigma}_i$ are operating in the
Gor'kov-Nambu ($N$) and Keldysh ($K$) subspaces, respectively. 
Additionally, $\mathrm{Tr}(\ldots)$ implies a trace over
all matrices and continuous indices while curly brackets $\{,\}$ stand
for the anticommutator. The action $\mathcal{S}_Q$ represents coupling
between the $\hat{Q}_{tt'}$-matrix field and the superconducting order
parameter field $\hat{\Delta}(t)$. The former is essentially a local
in space electronic Green's function in the island which is a matrix
in $K\otimes N\otimes T$ (time) spaces. The superconducting part of the
action $\mathcal{S}_\Delta$ stems from the Hubbrd-Stratonovich
decoupling of a bare four-fermion BCS interaction term, which is done
by introducing the $\hat{\Delta}$-field. The action is subject to the 
nonlinear constraint $\hat{Q}^2=1$. As usual for the Keldysh
theory~\cite{Review}, all fields come in doublets of classical and
quantum components. The former obey equations of motion, and the latter
serve to generate these equations along with the corresponding
stochastic noise terms. In particular,
\begin{equation}\label{Delta}
\hat{\Delta}=\hat{\Delta}^c\hat{\sigma}_0+\hat{\Delta}^q\hat{\sigma}_3,
\quad \hat{\Delta}^\alpha=\left(\begin{array}{cc}0 & \Delta^\alpha
\\ -\Delta^{*\alpha} & 0\end{array}\right)_N, \quad\vec{\Delta}=\left(\begin{array}{c}\Delta^c\\ \Delta^q\end{array}\right).
\end{equation}
The action $\mathcal{S}_T$ describes the coupling of the
$\hat{Q}$-matrix in the island to those in the leads,
\begin{equation}\label{Q-LR}
\hat{Q}^{[\chi]}_k=\left(\begin{array}{cc}\hat{h}_k &
-(1-\hat{h}_k)e^{i\chi_k\hat{\tau}_3} \\
-(1+\hat{h}_k)e^{-i\chi_k\hat{\tau}_3} &
-\hat{h}_k\end{array}\right)_K\hat{\tau}_3,
\end{equation}
where
$\hat{h}_k=h(\varepsilon-eV_k\hat{\tau}_3)=\tanh\big(\frac{\varepsilon-eV_k\hat{\tau}_3}{2T}\big)$
is the distribution function and $\chi_k$ is the counting field. The
latter is essentially a quantum component of the vector potential
which serves to generate observable current and its higher moments.
Finally, the $\mathcal{S}_H$ part of the action accounts for the dephasing term of
Cooper pairs due to the magnetic field. The action
$\mathcal{S}[Q,\Delta,\chi]$ defines the nonequilibrium partition
function via the functional integral over all possible realizations
of $\hat{Q}$ and $\hat{\Delta}$,
\begin{equation}\label{Z}
\mathcal{Z}(\chi)=\int\mathcal{D}[Q,\Delta]\exp(i\mathcal{S}[Q,\Delta,\chi]).
\end{equation}
Knowledge of $\mathcal{Z}$ yields all desired cumulants for
current fluctuation by the simple differentiation $\langle
I^n\rangle=(e/t_0)^{n}(-i\partial_\chi)^n\ln \mathcal{Z}(\chi)$.

\textit{Technicalities}.-- When computing the path integral in
Eq.~\eqref{Z} we need to identify such a configuration of the
$\hat{Q}$-matrix field that realizes the saddle point of the action
Eq.~\eqref{S-Q}. For this purpose one needs a parametrization of the
$\hat{Q}$-field which explicitly resolves the nonlinear constraint
$\hat{Q}^2=1$. We adopt the exponential parametrization
$\hat{Q}=e^{-i\hat{W}}\hat{Q}_0e^{i\hat{W}}$ with $\{\hat{W},\hat{Q}_0\}=0$, 
where the matrix multiplication in the time space is implicitly assumed. 
A new matrix field $\hat{W}_{tt'}$ accounts for the rapid fluctuations of
$\hat{Q}$ associated with the electronic degrees of freedom and is
to be integrated out, while $\hat{Q}_0$ is the stationary Green's
function. Minimizing the action Eq.~\eqref{S-Q} with respect to $\hat{W}$, one
finds the following saddle point equation for $\hat{Q}_0$,
\begin{equation}\label{Usadel}
\frac{\delta}{8\pi}\sum_{k}g_k\big[\hat{Q}_0,\hat{Q}^{[\chi]}_k\big]=
-\big\{\hat{\tau}_3\partial_t,\hat{Q}_0\big\}
+i\big[\hat{\Delta},\hat{Q}_0\big]+
\frac{1}{4\tau_H}\big[\hat{\tau}_3\hat{Q}_0\hat{\tau}_3,\hat{Q}_0\big]
\end{equation}
which is merely a zero-dimensional version of the Usadel
equation. In the stationary case and without superconducting
correlations, Eq.~\eqref{Usadel} is solved by such a $\hat{Q}_0$ that
nullifies the commutator in the left-hand side. This immediately
suggests a solution for $\hat{Q}_0$ that has to be chosen as a linear
combination of the $\hat{Q}$-matrices in the leads,
\begin{eqnarray}\label{Q-SP}
&&\hskip-.5cm
\hat{Q}_0=
\left(\alpha_1\hat{Q}^{[\chi]}_1+\alpha_2\hat{Q}^{[\chi]}_2\right)/\sqrt{N_\chi},\\
&&\hskip-.5cm
N_\chi=\frac{1}{(g_1+g_2)^2}
\left(g^2_1+g^2_2+g_1g_2\big\{\hat{Q}^{[\chi]}_1,\hat{Q}^{[\chi]}_2\big\}\right),\label{N}
\end{eqnarray}
where the factor $N_\chi$ ensures proper normalization. If one now uses
Eqs.~\eqref{Q-SP} and \eqref{N} back in the action Eq.~\eqref{S-Q}, then
the partition function of the normal double tunnel junction follows 
immediately, in agreement with Ref.~\cite{Dima-Review-1}.

The next step is to integrate out the fluctuations around the saddle
point. To this end, we linearize Eq.~\eqref{Usadel} with respect to
$\delta\hat{Q}_0=2i\hat{Q}_0\hat{W}$, and solve for the Cooperon matrix
field $\hat{W}$ to linear order in the superconducting field
$\hat{\Delta}$ by passing to Fourier space to invert the matrix
equation. The result is
\begin{equation}
\hat{W}_{\varepsilon\varepsilon'}=\frac{i}{2}
\frac{i\Gamma_\chi+(\varepsilon+\varepsilon')\hat{\tau}_3\hat{Q}_0}
{\Gamma^2_\chi+(\varepsilon+\varepsilon')^2}\big[\hat{\Delta},\hat{Q}_0\big],
\end{equation}
where $\Gamma_\chi=\tau^{-1}_{H}+E_{Th}\sqrt{N_\chi}$. Integrating
over $\hat{W}$ at the Gaussian level in Eq.~\eqref{Z}, $\int
D[W]\exp(i\mathcal{S}[W,\Delta,\chi])=\exp(i\mathcal{S}[\Delta,\chi])$,
one arrives at the effective action written in terms of
the superconducting order parameter only, 
\begin{subequations}
\begin{equation}\label{S-eff}
\mathcal{S}[\Delta,\chi]=\mathcal{S}_a[\Delta,\chi]+\mathcal{S}_b[\Delta,\chi]+\mathcal{S}_\Delta,
\end{equation}
\begin{equation}
\label{S-a}
\mathcal{S}_a=\frac{\pi}{2\delta}\mathrm{Tr}\left(\mathcal{\hat{C}}^a_{\varepsilon\varepsilon'}
(\hat{Q}_0(\varepsilon)\hat{\Delta}_{\varepsilon-\varepsilon'}\hat{\Delta}_{\varepsilon'-\varepsilon}
+\hat{\Delta}_{\varepsilon-\varepsilon'}\hat{\Delta}_{\varepsilon'-\varepsilon}\hat{Q}_0(\varepsilon'))\right),
\end{equation}
\begin{equation}
\label{S-b}
\mathcal{S}_b=\frac{\pi}{2\delta}\mathrm{Tr}\left(\mathcal{\hat{C}}^b_{\varepsilon\varepsilon'}
(\hat{\Delta}_{\varepsilon-\varepsilon'}\hat{\Delta}_{\varepsilon'-\varepsilon}
-\hat{\Delta}_{\varepsilon-\varepsilon'}\hat{Q}_0(\varepsilon')
\hat{\Delta}_{\varepsilon'-\varepsilon}\hat{Q}_0(\varepsilon))\right),
\end{equation}
\end{subequations}
where
$\hat{\mathcal{C}}^a_{\varepsilon\varepsilon'}=i(\varepsilon+\varepsilon')\hat{\tau}_3/
[(\varepsilon+\varepsilon')^2+\Gamma^2_\chi]$ and
$\hat{\mathcal{C}}^b_{\varepsilon\varepsilon'}=\Gamma_\chi\hat{\tau}_0/
[(\varepsilon+\varepsilon')^2+\Gamma^2_\chi]$ are Cooperon propagators. 
For technical
reasons of convenience, with the intermediate steps of the calculations
we choose to work in the gauge $\chi_1=\alpha_2\chi$ and
$\chi_2=-\alpha_1\chi$, and similarly for the voltages
$V_1=\alpha_2V$ and $V_2=-\alpha_1V$. Carrying out matrix products,
traces, and integrations with the help of Eqs.~\eqref{Delta},
\eqref{Q-LR}, and \eqref{Q-SP}, one eventually finds 
\begin{subequations}
\begin{equation}\label{S-GL}
\mathcal{S}[\Delta,\chi]=\frac{\pi}{4\delta}
\mathrm{Tr}\left(\vec{\Delta}^\dag_{-\omega}\hat{\Pi}_\omega(V,\Delta
T,\chi) \vec{\Delta}_\omega\right),
\end{equation}
\begin{equation}\label{L}
\hat{\Pi}_\omega=\left(\begin{array}{cc}-i\chi^2_v\phi &
\mathcal{E}-i\omega/T_c-\chi^2_v\eta
\\ \mathcal{E}+i\omega/T_c-\chi^2_v\eta & 2i \end{array}\right).
\end{equation}
\end{subequations}
Here we have used the notation $\chi^2_v=\chi^2-ieV\chi/T_c$.
Equation \eqref{S-GL} represents a time-dependent Ginzburg-Landau action
for nonequilibrium superconducting fluctuations. Off-diagonal
elements (retarded and advanced blocks) of the propagator matrix
$\hat{\Pi}_\omega$ carry information about the excitation spectrum
of fluctuations. The Keldysh block (quantum-quantum element of the
matrix $\propto \Delta^q\Delta^{*q}$) ensures
fluctuation-dissipation relations. The anomalous classical-classical
block accounts for the feedback of stochastic Langevin forces of
fluctuations due to the nonequilibrium quasiparticle background.

Performing the remaining path integration over $\Delta$ in Eq.~\eqref{Z}
with the action from Eq.~\eqref{S-GL}, one realizes that the
corresponding cumulant generation function for current fluctuations
is governed by the determinant of the Ginzburg-Landau propagator
[Eq.~\eqref{L}], namely,
$\ln\Delta\mathcal{Z}\propto\Det\hat{\Pi}_\omega$. 
We regularize $\Det\hat{\Pi}_\omega$ by normalizing it to itself taken
at zero counting field,
$\Det\hat{\Pi}_\omega\to\Det\hat{\Pi}_\omega(\chi)/\Det\hat{\Pi}_\omega(0)$,
and thereby find
\begin{equation}
\ln\Delta\mathcal{Z}=t_0\int\frac{d\omega}{2\pi}\ln\left[1-\frac{2\chi^2_v(\phi+\mathcal{E}\eta)}
{\mathcal{E}^2+\omega^2/T^2_c}\right]
\end{equation}
which upon final integration reduces to Eq.~\eqref{lnZ-SC}. From the
structure of the effective action~\eqref{S-eff}, and
also relying on previous studies~\cite{Ahmadian-PRB05,TDGL-6},
one can identify the essential physical processes affecting conductance
and noise. The first $S_a$ term in the effective action corresponds
to the density of states effect. Superconducting fluctuations
suppress the quasiparticle density of states near the Fermi level that
translate into a zero-bias conductance dip~\cite{Dos}. The second
$S_b$ term of the action corresponds to the inelastic Maki-Thompson
process~\cite{Maki-Thompson}, which can be thought of as resonant
electron scattering on the preformed Cooper pairs. The combined
effect of the two processes has a profound implication for the higher
cumulants of the current noise. The final remark is that the
Aslamazov-Larkin fluctuational correction~\cite{Aslamazov-Larkin} is
absent in our case since we are considering a zero-dimensional limit
while the latter relies essentially on the spatial gradients of the
superconducting order parameter.

\textit{Acknowledgments}.-- We would like to thank M.~Reznikov for
motivating this study and A.~Kamenev for a number of useful discussions. 
The work by D.~B. was supported by SFB/TR 12 of the Deutsche Forschungsgemeinschaft.
A.~L. acknowledges support from NSF Grant No. ECCS-1407875, and the hospitality of the Karlsruhe
Institute of Technology where this work was finalized.

\end{document}